\documentclass[aps,pre,reprint,superscriptaddress,nofootinbib,captions=justified]{revtex4-2}

\usepackage{amsmath,amssymb,graphicx,bm}
\usepackage{mathtools}
\usepackage{tikz}
\usepackage{txfonts}
\usepackage{hyperref}
\usepackage{CJK}
\CJKnospace
\usetikzlibrary{arrows.meta,positioning}

\tikzset{
  node/.style={circle, draw, thick, minimum size=8mm, inner sep=0pt,
               font=\small},
  edge/.style={line width=1.1pt},
}

\begin{document}
\begin{CJK*}{UTF8}{mj}

\title{Two variants of the friendship paradox:
The condition for inequality between them}

\author{Sang Hoon Lee (이상훈)}
\email{lshlj82@gnu.ac.kr}
\affiliation{Department of Physics, Gyeongsang National University, Jinju, 52828, Korea}
\affiliation{Research Institute of Natural Science, Gyeongsang National University, Jinju 52828, Korea}
\affiliation{Future Convergence Technology Research Institute, Gyeongsang National University, Jinju 52849, Korea}
\date{\today}

\begin{abstract}
\noindent
The friendship paradox---the observation that, on average, one’s friends have more friends than oneself---admits two common formulations depending on whether averaging is performed over edges or over nodes. These two definitions, the ``alter-based'' and ``ego-based'' means, are often treated as distinct but related quantities. This paper establishes their exact analytical relationship, showing that the difference between them is governed by the degree-degree covariance $\operatorname{Cov_\mathrm{n}}(k,k_{nn})$ normalized by the mean degree $\langle k\rangle$. Explicit examples demonstrate the three possible cases of positive, zero, and negative covariance, corresponding respectively to assortative, neutral, and disassortative mixing patterns. The derivation further connects the covariance form to the moment-based expression introduced by Kumar, Krackhardt, and Feld [\textit{Proc.\ Natl.\ Acad.\ Sci.} \textbf{121}, e2306412121 (2024)], which involves the $(-1)$st, $1$st, $2$nd, and $3$rd moments of the degree distribution. The two formulations are shown to be equivalent, as they should be: the moment-based representation expands the same structural dependence that the covariance form expresses in its most compact and interpretable form. The analysis thus unifies node-level and moment-level perspectives on the friendship paradox, offering both a pedagogically transparent derivation and a direct bridge to recent theoretical developments.
\end{abstract}

\maketitle

\section{Introduction}

The friendship paradox (FP), first articulated by Feld~\cite{Feld1991}, states that the average degree of an individual’s friends typically exceeds the individual’s own degree. This phenomenon arises because high-degree nodes are disproportionately represented when sampling along edges. In practice, two closely related but distinct average quantities are used. The \emph{alter-based} mean computes the expected degree encountered by following a uniformly random edge. The \emph{ego-based} mean first computes, for each node, the average degree of its neighbors and then averages that quantity over nodes. Both are natural formalizations of ``how popular one’s friends are on average,'' yet they need not coincide outside special cases such as regular or degree-uncorrelated networks. Beyond these average-based statements, recent work has further clarified how majority-type versions of the paradox---often phrased as `most people’s friends have more friends'---can be different from the classical FP. In a recent paper~\cite{FPP}, I analyze this so-called strong friendship paradox~\cite{ELee2019,Kooti2014,Wu2017,Lerman2016} and show how local degree structure can reverse or amplify the classical expectation.

Recent work by Kumar, Krackhardt, and Feld~\cite{Kumar2024} introduced a moment-based decomposition that expresses the difference between these two averages in terms of degree moments and an edge-level correlation coefficient, the \emph{inversity} $\rho$, which captures the correlation between origin degree and inverse destination degree across oriented edges. The present paper establishes a compact node-level covariance identity that expresses the difference between the alter-based and ego-based averages as a normalized covariance between degree and neighbor-averaged degree. The equivalence between the two formulations is shown explicitly, including a clear mapping between node-averaged quantities, edge-sampled expectations, and degree moments. This connection illuminates the roles of heterogeneity and mixing patterns~\cite{Newman2002} and provides a concise criterion for equality of the two variants.

\section{Tale of Two FP Types}

Consider an undirected, unweighted network of $N$ nodes and $M$ edges with
adjacency matrix $A=\{a_{ij}\}$, where $a_{ij}=1$ if nodes $i$ and $j$
are connected and $0$ otherwise. The degree of node $i$ is
\begin{equation}
k_i=\sum_j a_{ij} \,,
\end{equation}
and the average degree of the network is
\begin{equation}
\langle k\rangle_{\mathrm{n}} = \frac{1}{N}\sum_i k_i = \frac{2M}{N} \,,
\end{equation}
where $\langle x \rangle_{\mathrm{n}} = \sum_i x_i / N$ indicates the node-based average (equivalent to the expected number of $x$ for a uniformly sampled node) of the measure $x$ assigned on each node from now on.
Let $p(k)$ denote the degree distribution, so that
$\langle k^m\rangle_\mathrm{n} = \sum_k k^m p(k)$ represents the $m$th moment of
the degree; in general, the node-averaged value of a function of degree, $f(k)$, defined on a network is given by
\begin{equation}
\langle f(k) \rangle_\mathrm{n} = \sum_k f(k) p(k) \,.
\end{equation}

There are two ways to quantify the average number of friends of one’s friends,
depending on whether the averaging is performed over \emph{edges}
or over \emph{nodes}~\cite{Feld1991}:
\begin{enumerate}
\item[(1)] The \textbf{edge-averaged} (or \emph{alter-based}) mean degree~\cite{Kumar2024},
denoted by $\langle k_{\text{friend}}\rangle_\mathrm{n}$, measures the expected degree
of a node reached by following a randomly chosen edge. Since the
probability of selecting a node is proportional to its degree,
the corresponding distribution of neighbor degrees is
$q(k)\propto k\,p(k)$, properly normalized as
$q(k)=k\,p(k)/\langle k\rangle_\mathrm{n}$. The resulting mean is
\begin{equation}
\langle k_{\text{friend}}\rangle_\mathrm{n}
= \sum_k k\,q(k)
= \frac{\langle k^2\rangle_\mathrm{n}}{\langle k\rangle_\mathrm{n}} \,.
\end{equation}
This quantity is greater than or equal to $\langle k\rangle_\mathrm{n}$, yielding
the classical form of the FP:
$\langle k_{\text{friend}}\rangle_\mathrm{n} \ge \langle k\rangle_\mathrm{n}$.
\item[(2)] The \textbf{node-averaged} (or \emph{ego-based}) mean degree of
neighbors~\cite{Kumar2024}, denoted by $\langle k_{nn}\rangle_\mathrm{n}$, represents the average of the
mean neighbor degree $k_{nn}(i)$ over all nodes:
\begin{equation}
k_{nn}(i) = \frac{1}{k_i}\sum_j a_{ij}k_j
\Rightarrow
\langle k_{nn}\rangle_{\mathrm{n}} = \frac{1}{N}\sum_i k_{nn}(i) \,.
\end{equation}
Here, $k_{nn}(i)$ can vary across nodes depending on local connectivity
and degree--degree correlations. Its network average provides a second,
natural formulation of the same intuitive notion---how well connected
one’s friends are on average.
\end{enumerate}
In the terminology of Ref.~\cite{Kumar2024},
$\langle k_{\text{friend}}\rangle_\mathrm{n}$ corresponds to the
\emph{alter-based mean} and $\langle k_{nn}\rangle_\mathrm{n}$ to the
\emph{ego-based mean}. Although they coincide in regular or degree-uncorrelated networks,
their difference in general encodes the structural correlation between a
node’s degree and that of its neighbors~\cite{Newman2002}. The remainder of this paper
analyzes this difference and establishes its precise relationship to the
degree-degree covariance for each node and its nearest neighbors.

\section{Relation between the two FP types}

The two averages introduced above, $\langle k_{\text{friend}}\rangle_\mathrm{n}$ and
$\langle k_{nn}\rangle_\mathrm{n}$, arise from distinct averaging schemes yet are
closely related through the structure of the adjacency matrix.
The starting point is the identity
\begin{equation}
\sum_i k_i^2 = \sum_i k_i \sum_j a_{ij} \,,
\label{eq:sum_to_double}
\end{equation}
which expresses the sum of squared degrees as the sum of all degrees
counted once for every incident edge. Because the adjacency matrix
$a_{ij}$ is symmetric in an undirected network ($a_{ij}=a_{ji}$), the
same double sum can be viewed from either side of the edge:
\begin{equation}
\sum_i k_i^2
= \sum_i\sum_j a_{ij}k_i
= \sum_i\sum_j a_{ij}k_j \,.
\end{equation}
The two forms are equivalent by simply exchanging the dummy indices
$i\leftrightarrow j$ in the unrestricted double sum. Averaging both
sides over all nodes allows the two perspectives---``my degree counted
per friend'' and ``my friends' degrees counted per me''---to meet at the
same combinatorial total.
\medskip

Next, connect this double-sum identity to the local mean degree of
neighbors.  For each node $i$,
\begin{equation}
k_{nn}(i)=\frac{1}{k_i}\sum_j a_{ij}k_j \,.
\end{equation}
Multiplying both sides by $k_i$ and summing over all nodes gives
\begin{equation}
\sum_i k_i k_{nn}(i)
= \sum_i \sum_j a_{ij}k_j
= \sum_i k_i^2 \,.
\label{eq:sum_k2_equals_k_knn}
\end{equation}
Therefore,
\begin{equation}
\displaystyle
\sum_i k_i^2 = \sum_i k_i k_{nn}(i) \,.
\label{eq:k2_sum_and_k_and_knn_sum}
\end{equation}
Dividing both sides of Eq.~\eqref{eq:sum_k2_equals_k_knn} by $N$ and using
$\langle x\rangle_\mathrm{n} =\sum_i x_i / N$ yields
\begin{equation}
\langle k^2\rangle_\mathrm{n} = \langle k\,k_{nn}\rangle_\mathrm{n} \,.
\label{eq:k2_sum_and_k_and_knn}
\end{equation}
Substituting $\langle k_{\text{friend}}\rangle_\mathrm{n} =
\langle k^2\rangle_\mathrm{n} /\langle k\rangle_\mathrm{n}$ gives an exact relation
\begin{equation}
\langle k_{\text{friend}}\rangle_\mathrm{n}
= \frac{\langle k\,k_{nn}\rangle_\mathrm{n}}{\langle k\rangle_\mathrm{n}} \,.
\label{eq:kfriend_equals_kk_over_k}
\end{equation}
This identity shows that the difference between the alter-based and
ego-based means arises from the correlation between a node’s
degree and the mean degree of its neighbors. Subtracting
$\langle k_{nn}\rangle_\mathrm{n}$ from both sides of Eq.~\eqref{eq:kfriend_equals_kk_over_k} gives
\begin{align}
\langle k_{\text{friend}}\rangle_\mathrm{n}-\langle k_{nn}\rangle_\mathrm{n}
&=\frac{\langle k\,k_{nn}\rangle_\mathrm{n}
-\langle k\rangle_\mathrm{n} \langle k_{nn}\rangle_\mathrm{n}}
{\langle k\rangle_\mathrm{n}} \nonumber \\
&= \frac{1}{\langle k\rangle_\mathrm{n}}\operatorname{Cov_\mathrm{n}}(k,k_{nn}) \,,
\label{eq:cov_diff}
\end{align}
where the node-based covariance $\operatorname{Cov_\mathrm{n}}(k,k_{nn}) = \langle k\,k_{nn}\rangle_\mathrm{n}
-\langle k\rangle_\mathrm{n} \langle k_{nn}\rangle_\mathrm{n}$ indicates the covariance of the degree of a node and the average degree of its neighbors, a perfectly legitimate measure of degree assortativity~\cite{Newman2002}.

Equation~\eqref{eq:cov_diff} is the central analytical result.  It shows
that the two standard formulations of the FP coincide
when $\operatorname{Cov_\mathrm{n}}(k,k_{nn})=0$, that is, when node degrees and
their neighbors’ average degrees are statistically uncorrelated. A
positive covariance corresponds to assortative
mixing, yielding
$\langle k_{\text{friend}}\rangle_\mathrm{n} > \langle k_{nn}\rangle_\mathrm{n}$.  Conversely, a
negative covariance corresponds to
disassortative mixing, leading to
$\langle k_{\text{friend}}\rangle_\mathrm{n} < \langle k_{nn}\rangle_\mathrm{n}$.
Thus, the simple covariance term encapsulates all structural
correlations responsible for the inequality between the two versions of
the FP.

\section{Examples}

\subsection{$\langle k_{\textnormal{friend}}\rangle_\mathrm{n} = \langle k_{nn}\rangle_\mathrm{n}$ (neutral case, $\operatorname{Cov_\mathrm{n}}=0$)}

\begin{figure*}[t]
\begin{tabular}{lll}
(a) & (b) & (c) \\
{\centering
\begin{tikzpicture}[scale=1.0]
  \node[node] (A) at (0,1.2) {};
  \node[node] (B) at (1.8,1.2) {};
  \node[node] (C) at (0.9,0) {};
  \node[node] (D) at (3.0,0) {};
  \node[node] (E) at (4.2,0) {};
  \draw[edge] (A) -- (B);
  \draw[edge] (A) -- (C);
  \draw[edge] (B) -- (C);
  \draw[edge] (C) -- (D);
  \draw[edge] (D) -- (E);
\end{tikzpicture}
\par} &
{\centering
\begin{tikzpicture}[scale=1]
  \node[node] (A) at (0,1.6) {};
  \node[node] (B) at (-1.6,0.8) {};
  \node[node] (C) at (-1.2,-0.9) {};
  \node[node] (D) at (1.2,-0.9) {};
  \node[node] (E) at (1.6,0.8) {};
  \node[node] (F) at (0,0.25) {};
  \draw[edge] (A) -- (C);
  \draw[edge] (A) -- (D);
  \draw[edge] (A) -- (F);
  \draw[edge] (B) -- (E);
  \draw[edge] (B) -- (F);
  \draw[edge] (C) -- (D);
  \draw[edge] (C) -- (F);
\end{tikzpicture}
\par} &
{\centering
\begin{tikzpicture}[scale=1]
  \node[node] (C) at (0,0.2) {};
  \node[node] (A) at (0,1.65) {};
  \node[node] (B) at (-1.4,0.2) {};
  \node[node] (D) at (1.4,0.2) {};
  \node[node] (E) at (0,-1.4) {};
  \draw[edge] (C) -- (A);
  \draw[edge] (C) -- (B);
  \draw[edge] (C) -- (D);
  \draw[edge] (C) -- (E);
\end{tikzpicture}
\par}
\end{tabular}
\caption{(a) A simple five-node network where the alter-based and ego-based means coincide (neutral case). The network values are: $\langle k\rangle_\mathrm{n} = 2$, $\langle k_{\text{friend}}\rangle_\mathrm{n} = 11/5$, $\langle k_{nn}\rangle_\mathrm{n} = 11/5$, and $\operatorname{Cov_\mathrm{n}}(k,k_{nn}) = 0$. (b) A network with mild assortativity: high-degree nodes tend to connect to one another. The network values are: $\langle k\rangle_\mathrm{n} = 7/3$, $\langle k_{\text{friend}}\rangle_\mathrm{n} = 18/7$, $\langle k_{nn}\rangle_\mathrm{n} = 5/2$, and $\operatorname{Cov_\mathrm{n}}(k,k_{nn}) = 1/6$. (c) A star-like configuration exhibiting strong disassortativity. The network values are: $\langle k\rangle_\mathrm{n} = 8/5$, $\langle k_{\text{friend}}\rangle_\mathrm{n} = 5/2$, $\langle k_{nn}\rangle_\mathrm{n} = 17/5$, and $\operatorname{Cov_\mathrm{n}}(k,k_{nn}) = -36/25$.}
\label{fig}
\end{figure*}
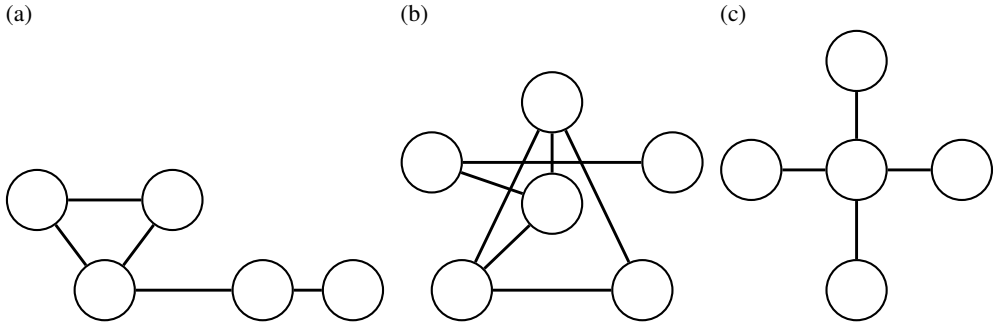

The example in Fig.~\ref{fig}(a) represents a \emph{neutral} structure: higher- and lower-degree
nodes coexist without systematic degree–degree correlation. As a result, the
FP appears identically under both the alter-based and ego-based
averages, illustrating $\langle k_{\text{friend}}\rangle_\mathrm{n}=\langle k_{nn}\rangle_\mathrm{n}$.

\subsection{$\langle k_{\textnormal{friend}}\rangle_\mathrm{n} > \langle k_{nn}\rangle_\mathrm{n}$ (assortative case, $\operatorname{Cov_\mathrm{n}}>0$)}

In the example in Fig.~\ref{fig}(b), the positive covariance indicates \emph{assortative mixing}, in which
high-degree nodes preferentially attach to other high-degree nodes.
Consequently, the edge-weighted (alter-based) average
$\langle k_{\text{friend}}\rangle_\mathrm{n}$ exceeds the node-weighted
(ego-based) average $\langle k_{nn}\rangle_\mathrm{n}$, consistent with
Eq.~\eqref{eq:cov_diff}:
\[
\langle k_{\text{friend}}\rangle_\mathrm{n}-\langle k_{nn}\rangle_\mathrm{n}
= \frac{1}{14}
= \frac{1}{\langle k\rangle_\mathrm{n}}\operatorname{Cov_\mathrm{n}}(k,k_{nn}) \,.
\]

\subsection{$\langle k_{\textnormal{friend}}\rangle_\mathrm{n} < \langle k_{nn}\rangle_\mathrm{n}$ (disassortative case, $\operatorname{Cov_\mathrm{n}}<0$)}

In the example in Fig.~\ref{fig}(c), the central node’s high degree paired with peripheral nodes of degree one
creates a clear disassortative structure. Because low-degree nodes dominate
the population, the node-averaged mean of neighbor degrees
($\langle k_{nn}\rangle_\mathrm{n}$) becomes larger than the edge-averaged mean,
yielding a negative covariance, the opposite limit of Fig.~\ref{fig}(b):
\[
\langle k_{\text{friend}}\rangle_\mathrm{n}-\langle k_{nn}\rangle_\mathrm{n}
= -\frac{9}{10}
= \frac{1}{\langle k\rangle}\operatorname{Cov_\mathrm{n}}(k,k_{nn}) \,.
\]

A simple example is the complete bipartite network composed of $n_1$ and $n_2$ nodes in the bipartite groups and all of the $n_1 n_2$ possible connections between the two groups exist, respectively, where
\begin{equation}
\langle k_{nn} \rangle_\mathrm{n} = \frac{n_1^2 + n_2^2}{n_1+n_2} = \frac{n_1^3 n_2 + n_1 n_2^3}{n_1^2 n_2 + n_1 n_2^2} = \frac{\langle k^3 \rangle_\mathrm{n}}{\langle k^2 \rangle_\mathrm{n}} \,,
\label{eq:bipartite}
\end{equation}
including the case in Fig.~\ref{fig}(c) with $n_1 = 1$ and $n_2 = 4$.

\section{Connecting the covariance and moment-based formulations}

The covariance form derived earlier,
\begin{equation}
\langle k_{\text{friend}}\rangle_\mathrm{n} - \langle k_{nn}\rangle_\mathrm{n} 
= \frac{1}{\langle k\rangle_\mathrm{n} }\operatorname{Cov_\mathrm{n}}(k,k_{nn}) \,,
\label{eq:cov_form}
\end{equation}
is mathematically equivalent to the moment-based expression
introduced in Ref.~\cite{Kumar2024}.
To make this equivalence explicit, let us reinterpret all relevant averages in terms
of degree moments and edge-sampled random variables.

\subsection{Recap of the degree-moment formalism of Ref.~\cite{Kumar2024}}

Let the network contain $N$ nodes and $M$ undirected edges.
Define the degree moments
\begin{equation}
\kappa_m = \frac{1}{N}\sum_{i=1}^{N} k_i^{m},
\qquad
\text{so that} \quad 2M = N\kappa_1 \,.
\label{eq:moments_def}
\end{equation}
Expectation values of a measure $y$ defined on a network taken over all of the edges $(i\!\to\! j)$
are denoted by 
\begin{equation}
\left\langle y (a_{ij}) \right\rangle_\mathrm{e} = \frac{1}{2M} \sum_{ij} a_{ij} y (a_{ij}) \,.
\end{equation}
Sampling a node uniformly from all edge endpoints yields the
\emph{edge-sampled} degree distribution (as is well known, the difference between the node-sampled and edge-sampled distributions is precisely the origin of the FP in the first place)
\begin{equation}
q(k) = \frac{k\,p(k)}{\langle k\rangle_\mathrm{n}}
      = \frac{k\,p(k)}{\kappa_1} \,,
\end{equation}
Therefore, the edge-averaged value of a function of degree, $g(k)$, defined on a network is given by
\begin{equation}
\langle g(k) \rangle_\mathrm{e} = \sum_k g(k)\,q(k) \,.
\label{eq:qk}
\end{equation}

Hence, the mean and the variance of the degrees of nodes attached to uniformly sampled edges are, in terms of the notations in Ref.~\cite{Kumar2024},
\begin{equation}
\mu_O = \langle k \rangle_\mathrm{e}
= \frac{\langle k^2\rangle_\mathrm{n}}{\langle k\rangle_\mathrm{n}}
= \frac{\kappa_2}{\kappa_1} = \langle k_{\text{friend}}\rangle_\mathrm{n} \,,
\label{eq:muO}
\end{equation}
and
\begin{equation}
\sigma_O^2
 = \langle k^2 \rangle_\mathrm{e} - \mu_O^2
 = \sum_k \frac{k^3 p(k)}{\kappa_1} - \left(\frac{\kappa_2}{\kappa_1}\right)^2 = \frac{\kappa_3}{\kappa_1}
   - \left(\frac{\kappa_2}{\kappa_1}\right)^2 \,,
\label{eq:sigmaO}
\end{equation}
respectively. On the other hand,
\begin{align}
\langle k_{nn}\rangle_\mathrm{n} = \frac{1}{N}\sum_i k_{nn}(i)
= \frac{1}{N}\sum_{ij} \frac{k_j}{k_i} a_{ij} \,.
\label{eq:E3a}
\end{align}
Since $2M = N\kappa_1$ oriented edges exist,
\begin{equation}
\langle k_{nn}\rangle_\mathrm{n} = \kappa_1 \left\langle \frac{D_D}{D_O} \right\rangle_\mathrm{e} \,,
\label{eq:E3}
\end{equation}
with the degree of the origin node $D_O=k_i$ and that of the destination node $D_D=k_j$.
Degrees at the two ends of an edge are correlated in general, so that
\begin{equation}
\left\langle D_D D_O^{-1} \right\rangle_\mathrm{e}
= \left\langle D_D\right\rangle_\mathrm{e} \left\langle D_O^{-1}\right\rangle_\mathrm{e}
 + \mathrm{Cov_\mathrm{e}}(D_D, D_O^{-1}) \,,
\label{eq:E4}
\end{equation}
which is precisely from the definition of covariance $\mathrm{Cov_\mathrm{e}}(D_D, D_O^{-1})$.
Since
\begin{equation}
\left\langle D_D\right\rangle_\mathrm{e} = \langle k \rangle_\mathrm{e} = \frac{\kappa_2}{\kappa_1} \,,
\label{eq:D_D_e}
\end{equation}
and 
\begin{equation}
\left\langle D_O^{-1}\right\rangle_\mathrm{e} = \left\langle \frac{1}{k} \right\rangle_\mathrm{e} = \sum_k \frac{(k/k) p(k)}{\kappa_1} = \frac{1}{\kappa_1} \,,
\label{eq:D_O_e}
\end{equation}
substituting Eq.~\eqref{eq:E4} into Eq.~\eqref{eq:E3} gives
\begin{equation}
\langle k_{nn}\rangle_\mathrm{n}
  = \kappa_2 / \kappa_1 
       + \kappa_1 \mathrm{Cov_\mathrm{e}}(D_D, D_O^{-1}) \,.
\label{eq:E5}
\end{equation}
Normalizing the covariance in Eq.~\eqref{eq:E5} defines the ``inversity'' $\rho = \mathrm{Corr_\mathrm{e}}\left( D_D,D_O^{-1} \right)$, which is the Pearson correlation coefficient between the degree of the origin node (``the origin degree'') and the reciprocal of the degree of the destination node (``the inverse destination degree''), introduced in Ref.~\cite{Kumar2024}:
\begin{equation}
\mathrm{Cov_\mathrm{e}}(D_D, D_O^{-1}) = \rho\,\sigma_O\,\sigma_{ID},
\label{eq:rho_def}
\end{equation}
with the edge-based variance of the inverse degree (ID)
\begin{align}
\begin{split}
\sigma_{ID}^2 & = \left\langle \left( \frac{1}{k} \right)^2 \right\rangle_\mathrm{e} - \left( \left\langle \frac{1}{k} \right\rangle_\mathrm{e} \right)^2 \\
 & = \sum_k \frac{k^{-1} p(k)}{\kappa_1} - \frac{1}{\kappa_1^2} \\
 & = \frac{\kappa_{-1}}{\kappa_1} - \frac{1}{\kappa_1^2} \,.
\end{split}
\end{align}
Thus, with the formula for $\sigma_O$ in Eq.~\eqref{eq:sigmaO}, we can write
\begin{align}
\begin{split}
\langle k_{nn}\rangle_\mathrm{n} & = \frac{\kappa_2}{\kappa_1} + \kappa_1 \rho \sqrt{\frac{\kappa_3}{\kappa_1} - \left( \frac{\kappa_2}{\kappa_1} \right)^2} \sqrt{\frac{\kappa_{-1}}{\kappa_1} - \frac{1}{\kappa_1^2}} \\
 & = \frac{\kappa_2}{\kappa_1}  + \rho \sqrt{\left(\frac{\kappa_1 \kappa_3 - \kappa_2^2}{\kappa_1} \right) \left( \kappa_{-1} - \kappa_1^{-1} \right) } \,.
\end{split}
\label{eq:k_nn_final}
\end{align}
Subtracting Eq.~\eqref{eq:muO} from Eq.~\eqref{eq:k_nn_final} from yields
\begin{equation}
\langle k_{nn}\rangle_\mathrm{n} - \langle k_{\text{friend}}\rangle_\mathrm{n} 
= 
\rho \sqrt{\left(\frac{\kappa_1 \kappa_3 - \kappa_2^2}{\kappa_1} \right) \left( \kappa_{-1} - \kappa_1^{-1} \right) 
} \,,
\label{eq:E9}
\end{equation}
which is exactly the same as in Ref.~\cite{Kumar2024}.

\subsection{Connecting the covariance forms and its implication}

The covariance in Eq.~\eqref{eq:cov_form} can be recast as
\begin{align}
\begin{split}
\operatorname{Cov_\mathrm{n}}(k,k_{nn}) & = \langle k k_{nn} \rangle_\mathrm{n} - \langle k \rangle_\mathrm{n} \langle k_{nn} \rangle_\mathrm{n} \\
 & = \kappa_2 - \kappa_1 \langle k_{nn} \rangle_\mathrm{n} \,,
\end{split}
\label{eq:Cov_formula}
\end{align}
and from Eq.~\eqref{eq:E5}, 
\begin{align}
\begin{split}
\operatorname{Cov_\mathrm{n}}(k,k_{nn}) & = -\kappa_1^2 \mathrm{Cov_\mathrm{e}} (D_D, D_O^{-1}) \,,
\end{split}
\label{eq:node_edge_cov_equiv}
\end{align}
which serves as the key link between the node-based and edge-based formulations. It demonstrates that the covariance between a node’s degree and the mean degree of its neighbors is directly proportional to the covariance between the degree at one end of an edge and the reciprocal degree at the other, both reflecting the same underlying degree-degree correlation~\cite{Newman2002}.

At this point, one can directly ``prove'' Eq.~\eqref{eq:node_edge_cov_equiv} by combining the relations discovered so far, starting from Eq.~\eqref{eq:Cov_formula}, and using Eq.~\eqref{eq:E3a}, Eq.~\eqref{eq:E4}, Eq.~\eqref{eq:D_D_e}, and Eq.~\eqref{eq:D_O_e} in the intermediate steps, i.e.,
\begin{align}
\begin{split}
\operatorname{Cov_\mathrm{n}}(k,k_{nn}) & = \kappa_2 - \frac{\kappa_1}{N} \sum_{ij} k_j \frac{1}{k_i} a_{ij} \\
 & = \kappa_2 - \frac{2M\kappa_1}{N} \left\langle D_D D_O^{-1} \right\rangle_\mathrm{e} \\
 & = \kappa_2 - \kappa_2 - \kappa_1^2 \mathrm{Cov_\mathrm{e}} (D_D, D_O^{-1}) \\
& = - \kappa_1^2 \mathrm{Cov_\mathrm{e}} (D_D, D_O^{-1}) \,.
\end{split}
\end{align}
Equation~\eqref{eq:node_edge_cov_equiv} is worth remembering because it connects two perspectives that are usually treated separately in the literature. On the left-hand side, the node-based covariance $\operatorname{Cov_\mathrm{n}}(k,k_{nn})$ quantifies the co-fluctuation between a node’s degree and the average degree of its neighbors—essentially a \emph{local} measure of assortativity. On the right-hand side, $\mathrm{Cov_\mathrm{e}}(D_D,D_O^{-1})$ describes how the degrees at the two ends of an edge co-vary across the entire \emph{edge ensemble}. The negative sign simply reflects that higher-degree origins correspond to smaller reciprocal factors $D_O^{-1}$, making the two covariances opposite in direction yet identical in magnitude up to the scaling factor $\kappa_1^2=(2M/N)^2$.

This mapping provides the crucial conceptual bridge between node-level and edge-level statistics. It reveals that what appears as a covariance between a node and its neighbors in the node-based ensemble can be equivalently interpreted as a covariance between coupled quantities across edges—one measuring the degree of the destination and the other the inverse degree of the origin. In networks with assortative mixing, high-degree nodes are likely to connect to other high-degree nodes, leading to a positive node-level covariance but a negative edge-level covariance because of the inverse transformation. Conversely, in disassortative networks, the node-level covariance becomes negative while the edge-level one becomes positive.

Therefore, Eq.~\eqref{eq:node_edge_cov_equiv} unifies the covariance-based and moment-based approaches by identifying them as two sides of the same structural dependency: one expressed through node-based sampling and the other through edge-based sampling. It also shows that both are naturally consistent with the classical degree-assortativity coefficient~\cite{Newman2002}.

\section{Discussion and conclusions}

We have demonstrated that the covariance form presented in this paper
and the moment–based expression of Ref.~\cite{Kumar2024} describe the
same underlying structural relation, albeit from two complementary
statistical viewpoints.
The \emph{inversity} parameter $\rho$ is defined as the Pearson correlation
on the oriented-edge ensemble between a node’s degree and the inverse degree
of its neighbor:
\begin{equation}
\rho = \mathrm{Corr_\mathrm{e}}(D_O, D_O^{-1})
      = \frac{\mathrm{Cov_\mathrm{e}}(D_O, D_O^{-1})}
             {\sigma_O \sigma_{ID}} \,.
\end{equation}
As noted in Ref.~\cite{Kumar2024}, $\rho$ cannot be expressed purely in terms of
the degree moments $\kappa_{-1}$, $\kappa_1$, $\kappa_2$, and $\kappa_3$,
because it depends on the \emph{joint} degree-degree distribution across edges.
The moment terms control the marginal degree heterogeneity,
whereas $\rho$ encodes the mixing pattern between adjacent nodes.
Consequently, the formulation in Ref.~\cite{Kumar2024} deliberately leaves $\rho$
as an irreducible correlation coefficient rather than attempting
a closed moment expansion.
By contrast, the covariance form in Eq.~\eqref{eq:cov_form}
\begin{equation}
\langle k_{\text{friend}}\rangle_\mathrm{n} - \langle k_{nn}\rangle_\mathrm{n}
   = \frac{1}{\langle k\rangle_\mathrm{n}}\operatorname{Cov_\mathrm{n}}(k,k_{nn}) \,,
\end{equation}
achieves an arguably simpler and more interpretable representation.
It summarizes the same degree-degree dependency at the node level,
requiring no additional joint-edge parameter.

Both approaches quantify how deviations from random connectivity
modulate the FP:
the moment form emphasizes the decomposition into degree moments
and edge-level correlation,
whereas the covariance form expresses the same dependence through
the direct co-fluctuation of a node’s degree and its neighbors' mean degree.
In this sense, Eq.~\eqref{eq:cov_form} may serve as the
most compact pedagogical statement of the phenomenon.
The equivalence between Eqs.~\eqref{eq:cov_form} and~\eqref{eq:E9}
shows that whether one adopts the node-level covariance picture
or the moment-based inversity framework,
both ultimately describe the same algebraic mechanism linking
degree heterogeneity, degree-degree correlation,
and the apparent inequality between one’s own connectivity
and that of one’s friends.
Finally, Eq.~\eqref{eq:node_edge_cov_equiv}, a by-product of the derivation, along with a crucial step Eq.~\eqref{eq:k2_sum_and_k_and_knn} to derive Eq.~\eqref{eq:cov_form}, may become a useful formula to calculate various degree-related quantities on networks. 

\acknowledgments

The author sincerely appreciates Hang-Hyun Jo (조항현) for verifying the covariance formula. The author also thanks Eun Lee (이은), with whom he initially discussed the problem and explored a plan to extend the analysis, as well as Mi Jin Lee (이미진) for organizing an informal workshop dedicated to the discussion of ``incomplete ideas'' where the topic was further developed. This research was supported by the National Research Foundation of Korea (NRF) under the grant RS-2021-NR061247.

\end{CJK*}


\begin{thebibliography}{99}
\bibitem{Feld1991}
\href{https://dx.doi.org/10.1086/229693}{S. L. Feld, Why your friends have more friends than you do, Am. J. Sociol. \textbf{96}, 1464 (1991)}.

\bibitem{FPP}
\href{https://link.springer.com/article/10.1007/s40042-025-01559-4}{S.\,H. Lee,
Friendship-paradox paradox: Do most people's friends really have more friends than they do?
J. Korean Phys. Soc. \textbf{88}, 890 (2026)}.

\bibitem{ELee2019}
\href{https://dx.doi.org/10.1103/PhysRevE.99.052302}{E. Lee, S. Lee, Y.-H. Eom, P. Holme, and H.-H. Jo,
Impact of perception models on friendship paradox and opinion formation,
Phys. Rev. E \textbf{99}, 052302 (2019)}.

\bibitem{Kooti2014}
\href{https://dx.doi.org/10.1609/icwsm.v8i1.14543}{F. Kooti, N.\,O. Hodas, and K. Lerman,
Network weirdness: Exploring the origins of network paradoxes,
in \textit{Proceedings of the International AAAI Conference on Weblogs and Social Media} \textbf{8}, 266 (2014)}.

\bibitem{Wu2017}
\href{https://dx.doi.org/10.1038/s41598-017-06042-0}{X.-Z. Wu, A.\,G. Percus, and K. Lerman,
Neighbor-neighbor correlations explain measurement bias in networks,
Sci. Rep. \textbf{7}, 5576 (2017)}.

\bibitem{Lerman2016}
\href{https://dx.doi.org/10.1371/journal.pone.0147617}{K. Lerman, X. Yan, and X.-Z. Wu,
The ``majority illusion'' in social networks,
PLOS ONE \textbf{11}, e0147617 (2016)}.

\bibitem{Kumar2024}
\href{https://dx.doi.org/10.1073/pnas.2306412121}{V. Kumar, D. Krackhardt, and S. Feld, On the friendship paradox and inversity: A network property with applications to privacy-sensitive network interventions, Proc. Natl. Acad. Sci. USA \textbf{121}, e2306412121 (2024)}.

\bibitem{Newman2002}
\href{https://dx.doi.org/10.1103/PhysRevLett.89.208701}{M.\,E.\,J. Newman, Assortative mixing in networks, Phys. Rev. Lett. \textbf{89}, 208701 (2002)}.

\end{thebibliography}
\end{document}